\documentclass[12pt]{article}         
\usepackage{epsfig}                   
\def\BB{Hei\-del\-berg--Mos\-cow $\beta\beta$ experiment}
\def\HdMo{Hei\-del\-berg--Mos\-cow $\beta\beta$ experiment}
\def\K40{$^{40}$K}

\def\KK{H.V.~Klapdor-Kleingrothaus}

\begin{document}
\title{\bf{First Results from the HDMS experiment in the Final Setup}}

\author{ 
H.V.~Klapdor-Kleingrothaus$^{\mbox{*}}$,  
A.~Dietz,
G. Heusser,\\
I.V. Krivosheina,
D. Mazza,
H. Strecker,
C. Tomei\\
}

\date{ }
\maketitle
\begin{center}
\begin{tabular}{c}                    
$^{\mbox{*}}$                   
Spokesman of the Collaboration\\     
\end{tabular}                         
\end{center}

\begin{abstract}
The Heidelberg Dark Matter Search (HDMS) is an experiment designed for
the search for WIMP dark matter.
It is using a special configuration of Ge detectors,
to efficiently reduce the background in the
low-energy region below 100 keV. 
After one year of running the HDMS detector prototype in the Gran Sasso 
Underground Laboratory, the inner crystal of the detector has been
replaced with a HPGe crystal of enriched $^{73}$Ge. The final setup
started data taking in Gran Sasso in August 2000.  
The performance and the first results of the measurement with the final setup are discussed.
\end{abstract}

\section{Introduction}                

Weakly Interacting Massive Particles (WIMPs) are leading candidates
for the dominant form of matter in our Galaxy.
These relic particles from an early phase of the Universe arise
independently from cosmological considerations in supersymmetric 
particle physics theories as neutralinos - the lightest supersymmetric
particles.

Direct WIMP detection experiments exploit the elastic WIMP scattering 
off nuclei in a terrestrial detector \cite{goodwitt}.
However, detecting WIMPs is not a simple task. 
Their interaction with matter is very feeble ($\sigma \leq \sigma_{weak}$) 
and predicted rates in supersymmetric models range from 10 to
10$^{-5}$ events per kilogram detector material and day
\cite{theo_rates,bedny1,bottino,bedny2,bedny3}. 
Moreover, for WIMP masses between a few GeV and 1 TeV, the energy
deposited by the recoil nucleus is less then 100 keV.
Thus, in order to be able to detect a WIMP, an experiment with 
a low-energy threshold and an extremely low radioactive background 
is required.
Since the reward would be no less than discovering the dark matter in
the Universe, a huge effort is put into direct detection experiments.
More than 20 experiments are running at present and even more are
planned for the future (for recent reviews see \cite{yorckz,lakla,yorckproc}).

\section{Description of the experiment}
In direct detection experiments looking for WIMPs one of the
main goals is the reduction of background events since the sensitivity of
experiments roughly scales with the obtained background level.

\begin{figure}[b!]
\begin{center}
\epsfxsize=60mm
\epsfbox{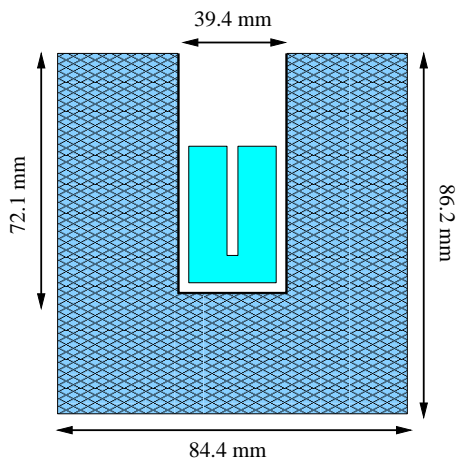}
\end{center}
\caption{\label{detindet_scheme}\small 
Schematic view of the HDMS detector configuration. All events which are seen in
both the inner and the outer Ge-detector can be considered as
background events, not resulting from a WIMP-nucleon recoil. 
}
\end{figure}

HDMS operates two ionization HPGe detectors in a unique configuration
\cite{prophdms}.
A small, p-type Ge crystal is surrounded by a well-type Ge crystal,
both being mounted into a common cryostat system (see
Figure~\ref{detindet_scheme} for a schematic view). 
Two effects are expected to reduce the background of the
inner target detector with respect to our best measurements with the
Heidelberg-Moscow experiment \cite{WIMPS}. 
First, the anticoincidence between the two detectors acts as an
effective suppression of multiple scattered photons. 
Second, we know that the main radioactive background of Ge detectors
comes from materials situated in the immediate vicinity of the
crystals. 
In the case of HDMS the inner detector is surrounded (apart from the
thin isolation) by a second Ge crystal - one of the radio-purest known
materials. 
In order to reduce the background with respect to the prototype
detector, the following changes has been made:

\begin{itemize}
\item The inner detector has been replaced by a crystal grown out of
HPGe material enriched in $^{73}$Ge. This has the effect that the mother 
isotope of cosmogenic $^{68}$Ge production, $^{70}$Ge 
($^{70}$Ge(n,t)$^{68}$Ge), is deenriched by
up to a factor of 50. Thus the decay of $^{68}$Ge will be suppressed by this
factor with respect to a natural HPGe crystal. The usage of enriched
$^{73}$Ge also allows the evaluation for spin dependent WIMP
nucleon crosssections.
\item The contacts of the HPGe crystals were pinched in order
to avoid the use of soldering tin inside the detector cap.
\item The crystal holder system has been replaced. The new material is from 
the same sample as the material used in the \HdMo, which is known to be
very clean. 
\end{itemize}
Some technical details are listed in Table \ref{hdmstech}.

\begin{table}
\caption{\label{hdmstech}}Technical data of the detector in the
final setup.\\[1ex]
\hspace*{.5cm}
\begin{tabular}{lrr}
\hline
Property&Inner Detector&Outer Detector\\
\hline
\hline
Crystal Type&p--type &n--type \\
Mass [g]&202&2111\\
Active Volume [cm$^3$]&37 &383 \\
Crystal diameter [mm]&35.2 &84.4 \\
Crystal length [mm]&40.3 &86.2 \\
Operation Bias [V]&+2500 &-1500 \\
Energy resolution FWHM (1332 keV) [keV] &1.87 &4.45 \\
Energy threshold [keV]&4.0 &7.5 \\
\hline
\end{tabular}

\end{table}

\begin{figure}[t!]
\begin{center}
\epsfxsize=70mm
\epsfbox{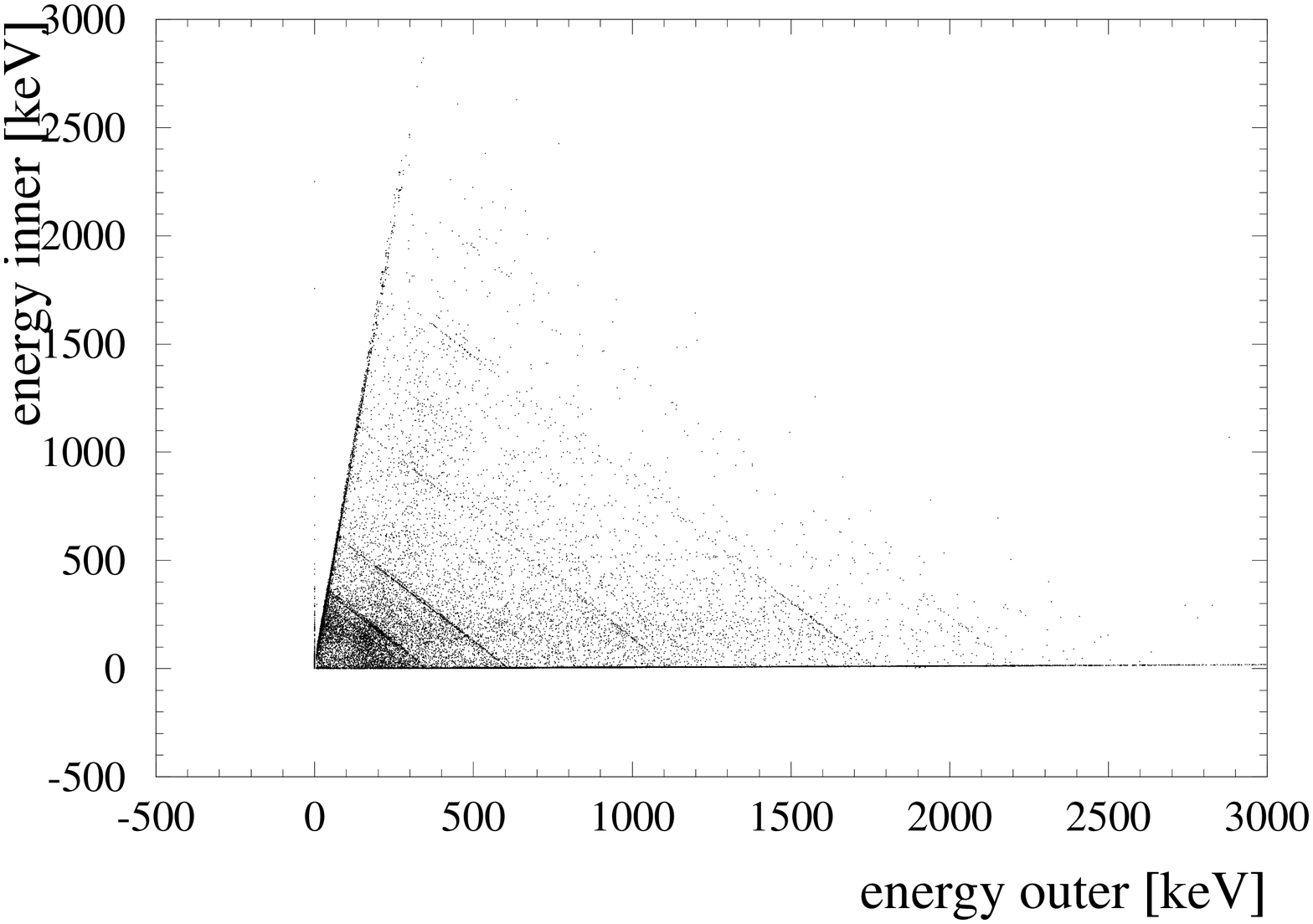}
\hfill
\epsfxsize=70mm
\epsfbox{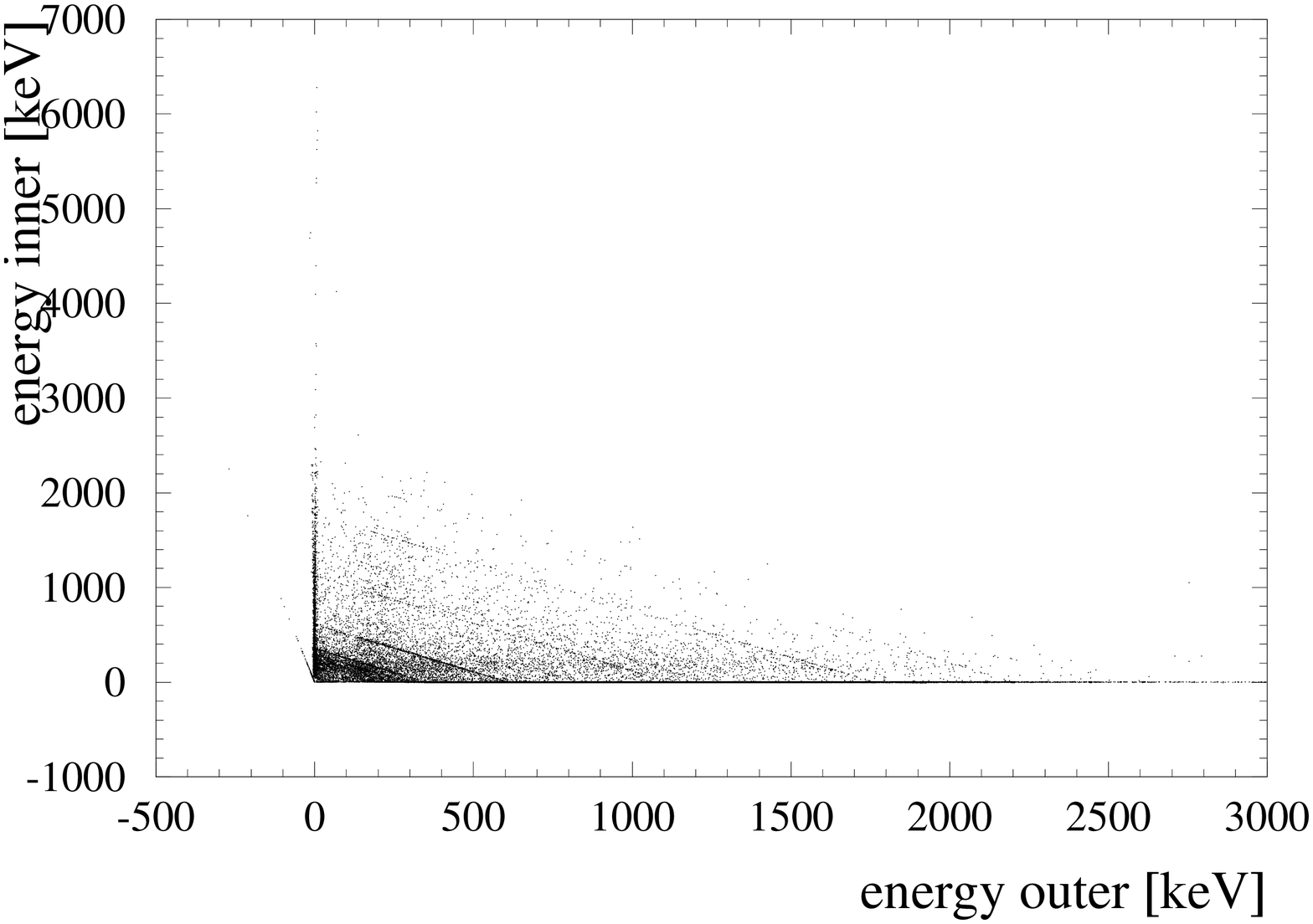}
\end{center}
\caption{\label{hdms_pickup}\small  Scatter plot of the measured data.
Each dot corresponds to one event. 
The y- and x-axis display the energy deposited in the inner- and outer detector, respectively.
Left: Before the correction for pick-up signals 
the zero energy axes have a non-zero slope.
Right: After the correction the zero energy axes correspond to the 
y- and x- axes.}
\end{figure}

\subsection{The anti-coincidence and the crosstalk}
Due to the special concentric design, the spatial separation between the two 
detectors is very small. This gives rise to pick-up signals. If one of the
detectors sees an event, a cross talk signal is induced in the other one.

Let $E_i$ and $E_o$ be the real deposited energies in the inner and
the outer detector, respectively. 
The cross talk effect causes pickup energies so that the measured
energies are:


\begin{equation}
E'_i=E_i+k_{io} \times E_o \quad \mbox{and} \quad E'_o=E_o+ k_{oi}\times E_i
\end{equation}

with the slopes $k_{io}$ and $k_{oi}$. 
The true energies deposited inside the detectors is given by the
relation

\begin{equation}
E_i=\frac{E'_i-k_{io}\times E'_o}{1-k_{io}\times k_{oi}} \quad
\mbox{and} \quad E_o=\frac{E'_o-k_{oi}\times E'_i}{1-k_{io}\times k_{oi}}
\end{equation}

Recording spectra of calibration sources with the list mode allows
to visualize this pick-up signal (see Fig. \ref{hdms_pickup}).
The anti-coincidence cut between the two detectors to
recognize multiple scattered background events can only be applied, 
if the cross talk is eliminated. 
This cut is made by defining all events as background events, 
in which an energy deposition
is seen in both detectors above the energy threshold of the proper detector.
 
It was shown that the cross talk is linear with energy and stable over time
and can be corrected for off-line \cite{yorck}.
Once the correction is made, the anti-coincidence can be applied.


\subsection{Measurements at the Gran Sasso underground laboratory}
The final setup of the HDMS was installed at the LNGS in August 2000,
the data used for the analysis are taken from February 2001 to September 2001
in order to let the cosmogenic isotopes decay.
The time stability of the energy resolution, threshold and calibration 
parameters (slope and intercept of energy calibration) is checked by
weekly measurements with a $^{152}$Ba-$^{228}$Th-source. 
This have been discussed elsewhere \cite{hdmsproto,lauradiss}.
The measured energy resolutions and thresholds of the detectors are listed
in Tab. \ref{hdmsres}. They correspond to standard values for detectors
of this size.

\begin{table}[t!]
\begin{centering}
\hspace*{-0.3cm}
\begin{tabular}{lcccc}
\hline
Property  &Inner Detector &Outer Detector\\
          & \multicolumn{2}{c}{final setup}\\
\hline
\hline
Threshold &(4.0$\pm0.2$)keV&(7.5$\pm0.2$)keV\\
\hline
Energy res.\\
at zero keV&(0.76$\pm0.05$)~keV&(2.82$\pm0.06$)~keV\\
extrapolated\\ \hline
at zero keV&(0.83$\pm0.01$)keV&(2.91$\pm0.04$) keV\\
after correction\\ \hline
at 81 keV  & (0.95$\pm0.03$) keV\\        \hline
at 344 keV &               &(3.03$\pm0.03$)keV\\ \hline
at 1408 keV&               &(4.46$\pm0.02$)keV\\
\hline 

\end{tabular}
\caption{\label{hdmsres}\small  Energy resolutions for different energies
and thresholds of the current final detector confi\-guration.}
\end{centering}
\end{table}

\begin{figure}[t!]
\begin{centering}
\epsfysize=90mm
\epsfbox{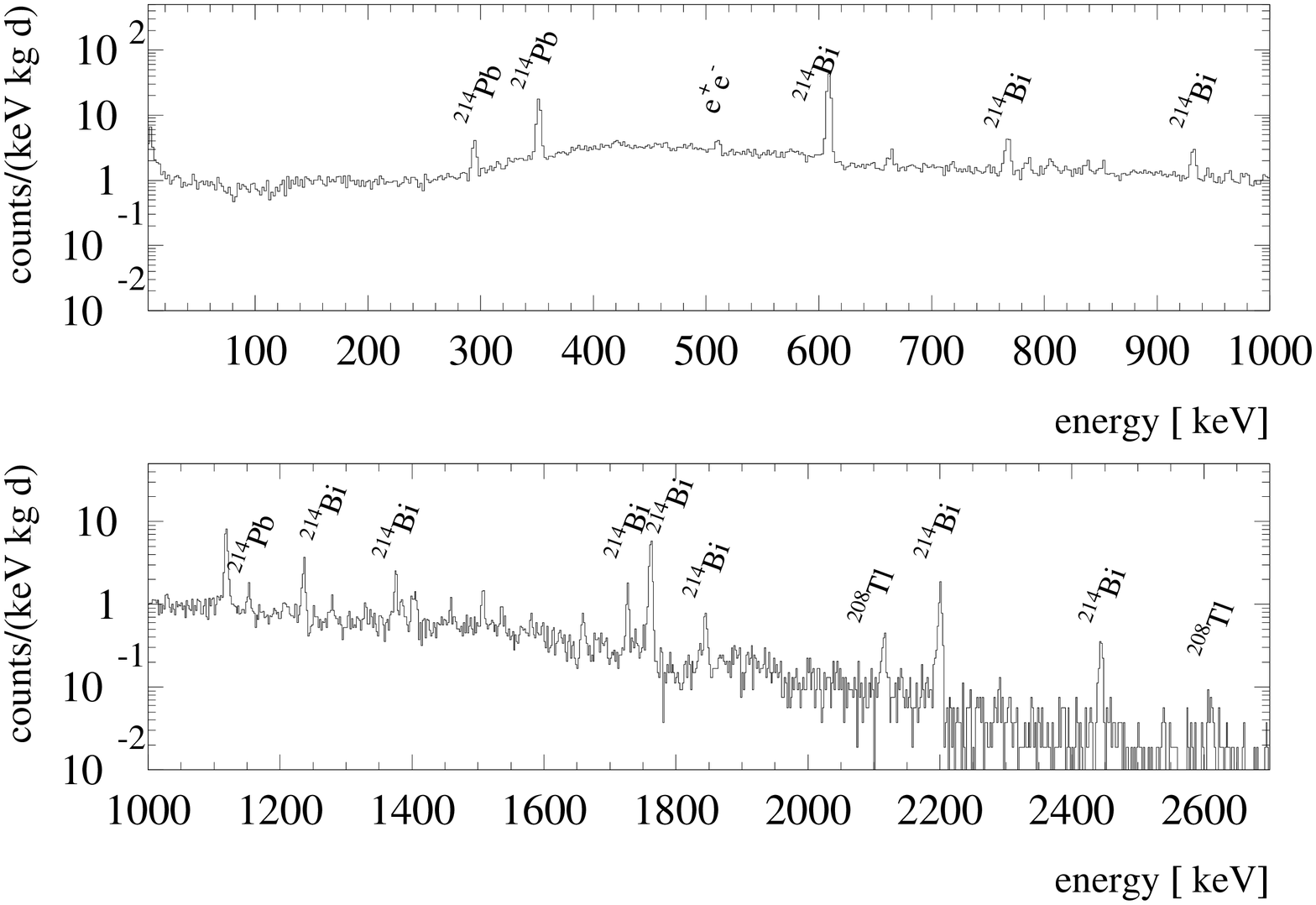}
\epsfysize=90mm
\epsfbox{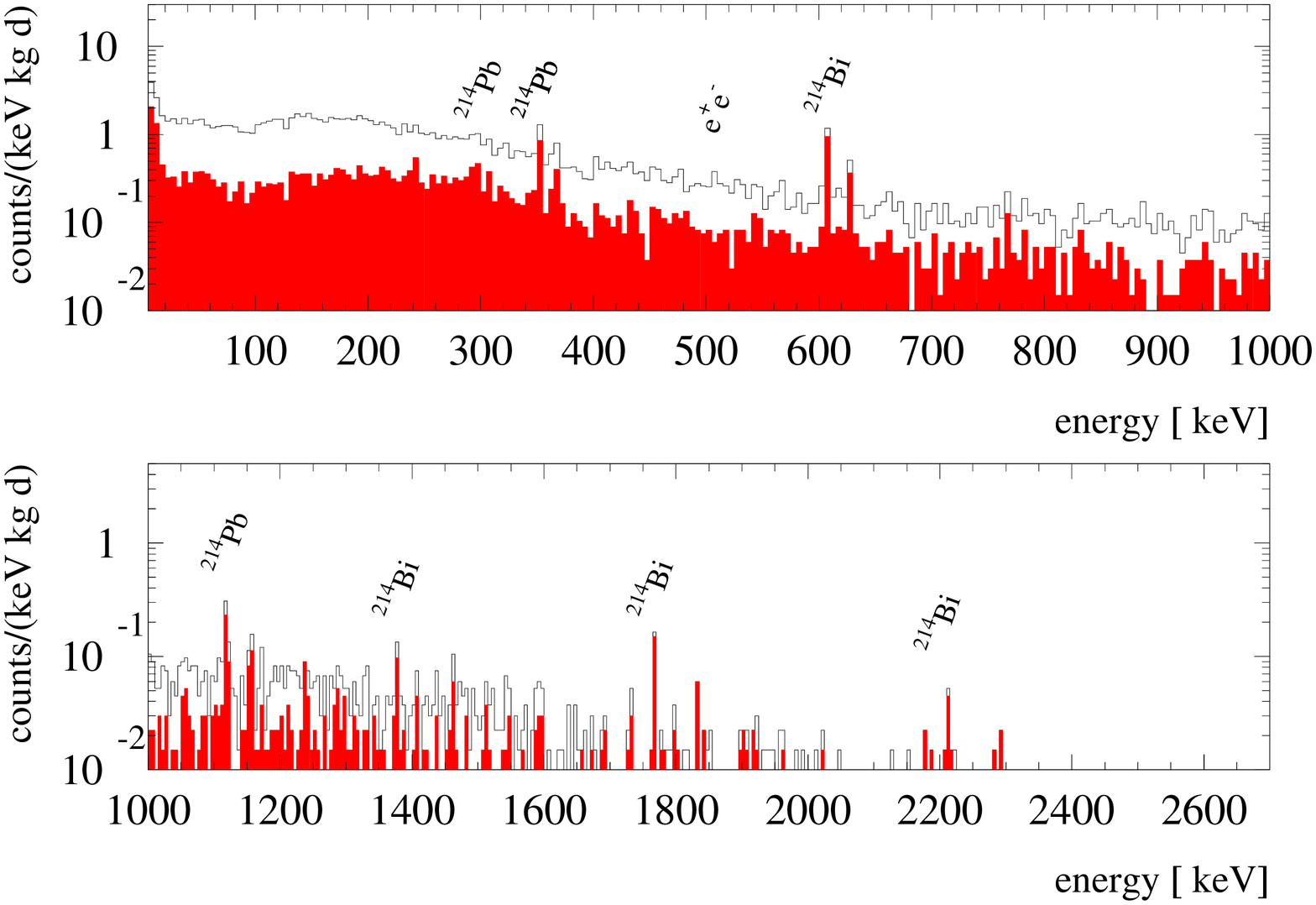}
\caption{\label{hdmsspecsfull}\small Spectra of the HDMS detectors of the
  final setup after a total measuring time of 132.4 days. Upper
  figure: outer detector; lower figure: inner detector; The
  open histogram denotes the overall spectrum, the filled histogram
  corresponds to the spectrum after the anti-coincidence cut with the
  outer detector. The most prominent lines are labeled.} 
\end{centering}
\end{figure}

The individual typical duration of a run was about 23 hours. 
The experiment was stopped daily and the most important detector
parameters like leakage current and mean count rates were checked. 
No substantial fluctuations were recorded. 

After the individual runs were calibrated and corrected for the
crosstalk, they were added to provide the sum spectra.
From the sum spectrum of the inner detector (after the anti-coincidence cut)
the limits on WIMP dark matter can be extracted.

\begin{figure}[th!]
\epsfxsize=10cm
\hspace*{2.5cm}
\epsfbox{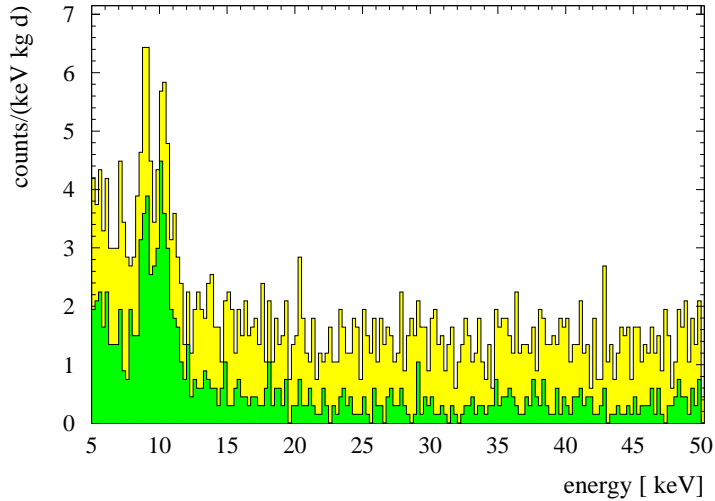}
\caption{\small Low-energy spectrum of the inner, enriched Ge detector before
  and after (open and solid histograms, respectively) the anti
  coincidence is applied with the outer Ge detector. The internal,
  low-energy x rays are 
  not removed by the anti-coincidence. \label{specilow}}
\end{figure}

Although the statistics of the inner detector is low, the 
sum-spectrum (see Fig. \ref{hdmsspecsfull}) show some lines of
isotopes from the U/Th natural decay chains as well as some X-ray
lines which are dominating the region below 10\,keV(see Fig. \ref{specilow}).
The most obvious structure in the low energy region is a peak at
10.37~keV resulting from the decay of $^{68}$Ge and a peak around
9~keV, which could arise from $^{65}$Zn. 
Note that for the low energy region (Fig. \ref{specilow})
there are no more indications for a contamination with $^{210}$Pb
in comparison with the results from the prototype-detector
\cite{hdmsproto}.
Also the structure at 32.5~keV, which is meanwhile understood (see \cite{dm2000}),
vanished completely.

If the anti-coincidence is evaluated in the energy region between
50\,keV and 100\,keV, the background reduction factor is 4.5. 
The counting rate after the anti-coincidence in this energy region is 
0.27\,events/(kg\,d\,keV) (from February 2001), thus comparable to the
value measured in the Heidelberg-Moscow experiment with the enriched
detector ANG2 \cite{WIMPS}.     
In the energy region between 11\,keV and 40\,keV the background index is 
0.43\,events/(kg\,d\,keV.).

\subsection{Dark Matter limits}
Since many cosmogenic isotopes have half-lifes below 300 days, typically the
count rate in low-level detectors decreases considerably after one year of
storage underground. 

For this reason only the last 132.4 days, corresponding to 26.74 kg d 
of measurement for the inner detector were used for the evaluation of
the final HDMS setup data.
The procedure of extracting limits on WIMP dark matter from the obtained 
spectrum follows the method described in \cite{WIMPS}.

\begin{figure}[t]
\epsfxsize=120mm
\begin{center}
\epsfbox{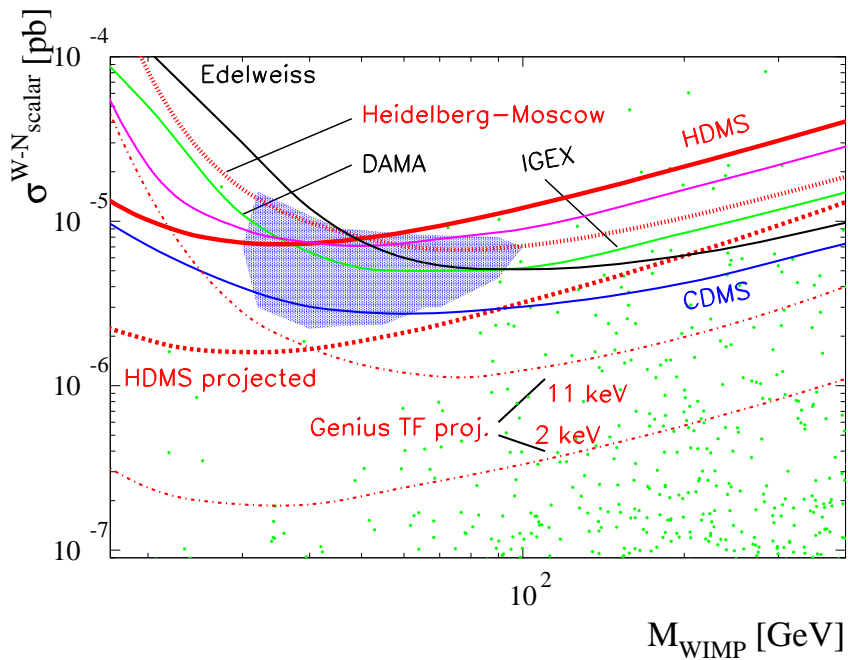}
\end{center}
\caption{\label{hdmsproto} 
\small Exclusion plot for the presently most sensitive WIMP dark
matter direct search experiments.  
The shaded area represents the 3$\sigma$ region allowed by the DAMA
experiment \cite{damacont}.
Already now, some parts of this region can be tested by the HDMS experiment.
Also shown are the present limits from the Heidelberg-Moscow
experiment \cite{WIMPS}, the DAMA experiment 
\cite{damaexcl}, the CDMS experiment \cite{cdmsexcl}, the IGEX
experiment \cite{igexexcl}, the Edelweiss experiment \cite{edwexcl}
and future expectations for HDMS and GENIUS-TF.
}
\end{figure}

The resulting upper limit is shown in the exclusion plot in figure 
\ref{hdmsproto}. 
Already now the limit given by the HDMS experiment is slightly better
in the low energy range
than the limit given by the \BB{} for low WIMP masses. 
This due to the fact that the energy threshold of 5~keV has been
obtained for this measurement (compare to 9~keV threshold of the \BB ~
\cite{WIMPS}). 

\section{Conclusion}
The detectors for the final HDMS setup, constructed with some
improvements made to reduce the background and including an inner
detector from enriched $^{73}$Ge, were installed in the LNGS 
in August 2000.
It took data for 132.4 days corresponding to 26.74 kg days.
The offline correction for the crosstalk has been applied to the data
and a sum spectrum was created, where most of the background sources
were identified. 
The background reduction factor in the inner detector through
anti-coincidence is about 4.5. 
The background in the low-energy region of the inner
detector (with exception of the region still dominated by cosmogenic
activities) is already comparable to the most sensitive dark matter
search experiments. 
It will be further decreasing due to the decay of cosmogenic isotopes, 
and with an improved smaller threshold HDMS should be able to test the
DAMA evidence region within about three years.

\end{document}